Reproducibility via Crowdsourced Reverse Engineering: A Neural Network Case Study With DeepMind's Alpha Zero

Dustin Tanksley and Donald C. Wunsch II

Applied Computational Intelligence Laboratory

Missouri University of Science and Technology

**Abstract**

The reproducibility of scientific findings are an important hallmark of quality and integrity in research. The scientific method requires hypotheses to be subjected to the most crucial tests, and for the results to be consistent across independent trials. Therefore, a publication is expected to provide sufficient information for an objective evaluation of its methods and claims. This is particularly true for research supported by public funds, where transparency of findings are a form of return on public investment. Unfortunately, many publications fall short of this mark for various reasons, including unavoidable ones such as intellectual property protection and national security of the entity creating those findings. This is a particularly important and documented problem in medical research, and in machine learning. Fortunately for those seeking to overcome these difficulties, the internet makes it easier to share experiments, and allows for crowd-sourced reverse engineering. A case study of this capability in neural networks research is presented in this paper. The significant success of reverse-engineering the important accomplishments of DeepMind's Alpha Zero exemplifies the leverage that can be achieved by a concerted effort to reproduce results.

## 1. Introduction

1.1 Reproducibility

Concerns about reproducibility have received increased visibility for several reasons, particularly since the Internet offers the opportunity for more data sharing than was available to previous generations. Furthermore, several prominent studies have shown that many reports in highly-reputed journals were not reproducible. (Baker, 2016) surveyed 1,500 scientists and found "More than 70% of researchers have tried and failed to reproduce another scientist's experiments, and more than half have failed to reproduce their own experiments" and that "52% of those surveyed agree that there is a significant 'crisis' of reproducibility". Therefore, many top journals are demanding improvements in the description of research and sharing of data to enhance reproducibility. (Greene, Garmire, Gilbert, Ritchie, & Hunter, 2017) states that "Journals, although they conduct peer review, do not validate each experimental result or claim.", and goes on to say

> "The act of rigorous secondary data analysis is critical for maintaining the accuracy and efficiency of scientific discovery. As scientists, we make predictions, perform experiments and generate data to test those predictions. When we ask rigorous questions, we obtain more accurate findings that can prevent harm."

(Committee on Reproducibility and Replicability in Science et al., 2019) notes how difficult reproducibility is, stating

> "When results are produced by complex computational processes using large volumes of data, the traditional methods section of a scientific paper is insufficient to convey the necessary information for others to reproduce the results. Additional information related to data, code, models, and computational analysis is needed."

However, the desire for reproducibility can conflict with legitimate interests of the researchers who provide the initial results. The pros and cons of sharing results have been vigorously debated in the medical literature. But this is even more salient when commercial researchers have proprietary responsibilities that may preclude sufficient disclosure to allow reproducibility. A prominent example of this in neural networks are the reports of the breakthrough accomplishments of DeepMind's AlphaGo series of systems. There is little doubt of the financial value of their neural network-related intellectual property. As of this writing, in three years since the victory, Google grew by approximately 70% in market value, a gain of over $350 billion. Of course, not all of that is attributable to the victory, but it is still clear that the company's reputation in neural-network based reinforcement learning is a significant part of its perceived value. The victory not only generated positive media coverage and bolstered the prospects of AI-enabled products, but even greatly influenced national budgets with South Korea alone increasing AI spending by 860 million (Zastrow, 2016), and China trying to draw equal to the US by 2020, and expanding its AI industry to $150 billion by 2030 ("Beijing Wants A.I. to Be Made in China by 2030— The New York Times," 2017; "China wants to be a $150 billion world leader in AI by 2030," 2017)

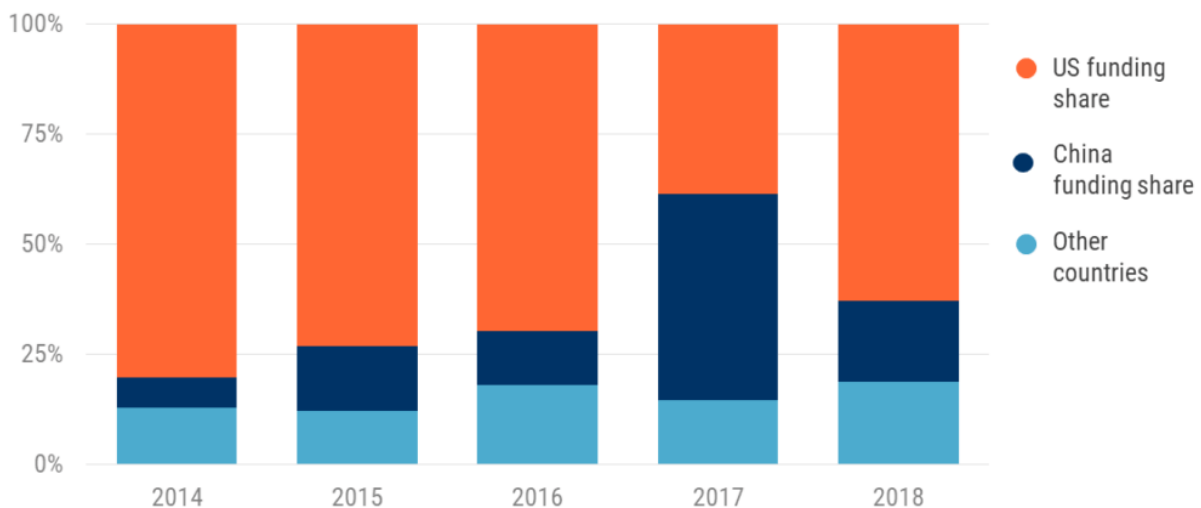

Figure 1 China drastically increases AI spending following Alpha Go's Victory ("China Is Starting To Edge Out The US In AI Investment—CB Insights Research," 2019)

Although DeepMind was justified in retaining key information, reproducibility of these important results was initially limited. Only the generalized search and training algorithms were provided, not the full source code or neural network architecture that was used. Trying to directly reproduce results using machine learning provides a unique challenge, as it is very costly due to the computing power needed to validate a single trial, much less actually finding the correct architecture and hyperparameters. DeepMind clearly spent many millions on computing resources alone to achieve these results. The labor hours and training time of failed models further increased the cost. Furthermore, previous progress in Go was among the impressive accomplishments that led to the US $500 million equivalent purchase of DeepMind by Google in 2014. The extraordinary economic impact of this research discussed above justifies these expenditures. It therefore is not surprising that details that would enhance reproducibility were withheld. If source code and final training weights were provided however, the cost to replicate results would drop by several orders of magnitude. In lieu of that, reverse-engineering for reproducibility may be the only viable option. This is difficult, but not nearly as difficult as developing the contribution in the first place. This example clearly illustrates the overwhelming importance of reverse-engineering for reproducibility. As a case study, it is highly instructive for future efforts.

## 2. Development of Go-playing Systems and Timeline of AlphaGo

Most Go researchers outside the DeepMind group did not expect a champion-level system in the near term, for good reason. Go is much harder than any of the games solved by previous AI systems. The game tree of Go is estimated to be between $10^{575}$ and $10^{620}$, hundreds of orders of magnitude greater than the game tree for Chess estimated around $10^{120}$ (L. Victor Allis, 1994; L.V. Allis, Herik, & Herschberg, 1990). Fundamental research has been conducted over a period of decades. (Silver, Sutton, & Müller, 2007) notes that it speculated that Go "requires an altogether different approach to other games." (Krikke, 2007) even foresaw that "methods which are successful at determining the value of Go positions might prove useful for image processing, as the analysis of Go positions is a very visual task" . Alpha Go used these, and many other insights that had built up over time, notably training a neural network to screen for a smaller subset of moves (Zaman & Wunsch, 1999), training via reinforcement learning in self-play (Zaman, Prokhorov, & Wunsch, 1997) and Monte-Carlo-based methods (D. Silver, 2009).

However, none of these systems were scaled up to nearly the level of AlphaGo. The human design effort and computational resources for training the system were unprecedented. DeepMind was given access to thousands of Google's proprietary TPUs, new hardware specifically designed for machine learning calculations, and trained for months. Even in the context of the extensive resources invested, the accomplishment was a major milestone in the history of neural networks and AI. Indeed, public and private investment in the field has dramatically increased since this accomplishment.

The same researchers announced an even more significant result just one year later (Silver, Schrittwieser, et al., 2017). They developed a new system, named AlphaGo Zero, that learned without historical game data, but purely by self-play. Furthermore, it performed far better than the original system, renamed AlphaGo Lee (in honor of Sedol), and required only 40 days of training (albeit on state-of-the-art hardware) compared to months for the previous research.

The system was then generalized to one that could learn arbitrary turn-based games, called Alpha Zero, with convincing performance learning from purely win-loss data on several games, notably chess, which only required 9 hours of training (Silver, Hubert, et al., 2017). It is notable that in 1000 games against the strongest rule based chess engine, Stockfish, Alpha Zero won 155 games, and lost only 6. This result was far beyond what many expected, as computers had already dominated chess for decades. Even Demis Hassabis, CEO of DeepMind noted in (Sadler & Regan, 2019) "it was far from clear that a program of this type could possibly hope to compete with the specialist handcrafted chess engines that had decades of cumulative effort spent on them from some of the best computer scientists and chess grandmasters in the world." If these results could be independently reproduced, it would confirm the importance of the tools described in their research.

### 3. Timeline of Reverse-Engineered Reproducibility Studies of Alpha Zero

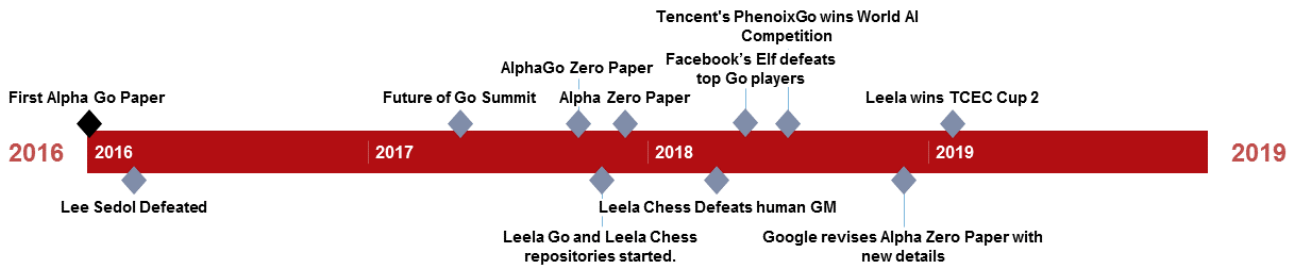

*Figure 2 A timeline of Alpha Go publications, and major events that happened in between, including recent replication efforts in Go and Chess.*

When DeepMind published their first paper on AlphaGo, interest began to build over the new method, which quickly became international news with the defeat of Lee Sedol. Despite this, replications efforts really went nowhere outside of promises for more investments into AI research. This initial version had used hundreds of thousands of master level games to train itself before being self-training, and direct access to this volume of data was difficult to get. This was a major roadblock until AlphaGo Zero was introduced. AlphaGo Zero managed to learn the entire game thru only self-play, meaning no data repository was needed to train the agent. Quickly group efforts were established to combine computing resources to replicate the results published, and this only spread more rapidly when DeepMind shared that the methods had been applied to other games, chess and shoji, with successful results.

The results in chess provided a unique opportunity for reverse-engineered reproducibility, as rule based game engines had already reached the pinnacle of chess play, where Go systems had no reliable comparison tool other than top ranked humans who couldn't compete with Alpha Go anymore. As well, DeepMind claimed only 9 hours of training time for chess, versus the 40 days they had used for AlphaGo Zero. Despite the fact that training would still be much slower due to the not having the resources to dedicate that DeepMind had, it would be much faster than Go.

Although DeepMind did not provide source code or neural network weights, they did provide enough details, that, with some assumptions, a similar system could be constructed. This was done and released in late 2017 ("LCZero," n.d.) and the interested community was invited to contribute to the code and the computational effort of training the system, named Leela Chess. Many community members suggested improvements for the system on a public forum, while the GitHub project was slowly expanded to fix issues with the architecture being used and upgraded for more efficient computations. Interested people could download a client, which would run simulations on their computer, and the results would be sent to a central repository where after several tens of thousands of games, a new network would be trained. If the new network performed better than the old one, clients would be updated with the new network and continue running simulations. Around this time Google Colabs introduced free GPU resources as well, allowing people to use their free computing power from Colabs to help contribute if they didn't have a powerful computing system (Bob23, 2018a; "Colaboratory – Google," n.d.).

Training on diverse systems of much lower capacity took longer, several months. But the system gradually moved from slow, weak play to grandmaster level, beating its first human GM in April 2018 ("GM Andrew Tang Defends Humanity Against Leela Chess Zero," 2018). Soon it began competing against other top engines in the TCEC (Top Chess Engine Championship), and in February 2019 managed to win the TCEC cup 2 playing on par with the best rule-based chess agents (Bob23, 2019b). The architecture, software and weights are in the public domain and this independent confirmation of the findings confirm the suitability and efficacy of the techniques proffered by DeepMind researchers. In addition to documenting the important independent confirmation of this achievement, this paper is also documenting the value of crowdsourced reverse engineering as a resource for independent reproducibility. Although more difficult, this independent confirmation is even more important than it would have been with shared code, because it leaves no question of the efficacy of the original contribution.

While the results in chess are very interesting, there are also plenty of results in Go as well. The Leela Zero project was the original inspiration for the Leela Chess Zero project and has steadily become stronger. Facebook has released an open source version of their Go AI named Elf, which managed to defeat top human Go players 20-0. One of the leading AI companies in China, Tencent, has also released an open source version of its Phoenix Go program, which won the 2018 World AI competition. This pursuit of independent, alternative implementations of important research ideas is a laudable trend that should be encouraged in the field.

### 4. Discussion: Implications and Resources

The challenge of reproducibility is increased in neural networks research due to the resource demands of many simulations. The software development effort can also be significant. Open source software and crowdsourcing of computational resources can mitigate these challenges. While in many cases this may become intractable, the approach should not be overlooked. Both high-performance computing and software neural network model development have steadily lowered barriers to entry, so that an approach that may be impractical in one year might subsequently become workable. The approaches described in this paper may therefore become more common. Policies could even be developed to encourage this type of independent verification (or refutation) of important research results.

## 5. Conclusion

Given the accelerating scale of investments in neural networks in particular, and research in general, by diverse stakeholders, many important contributions will not be described in a manner to allow immediate confirmation or refutation by the scientific community. However, this must not deter scientists from assessing reproducibility of the most important scientific results, with or without the help of the original researchers. Indeed, confirmation of results by researchers who are not among the original collaborators lends additional credibility to research results. This paper presented a case study of precisely such an approach to DeepMind's Alpha Zero design. This approach has provided a much broader community of researchers with the opportunity to confirm or refute results.

**Acknowledgements**

Partial support for this research was received from the Missouri University of Science and Technology Intelligent Systems Center, The Missouri University of Science and Technology Chancellor's Distinguished Fellowship, the U.S. Department of Education Graduate Assistance in Areas of National Need program, the Mary K. Finley Missouri Endowment, the National Science Foundation, the Lifelong Learning Machines program from DARPA/Microsystems Technology Office, and the Army Research Laboratory (ARL); and it was accomplished under Cooperative Agreement Number W911NF-18-2-0260.

The views and conclusions contained in this document are those of the authors and should not be interpreted as representing the official policies, either expressed or implied, of the Army Research Laboratory or the U.S. Government. The U.S. Government is authorized to reproduce and distribute reprints for Government purposes notwithstanding any copyright notation herein.